\documentclass[copyright,creativecommons]{eptcs}

\pdfoutput=1

\pdfmapfile{+newtx.map}

\usepackage[natbib=false,typeface=Times]{wjtools}

\hypersetup{hidelinks}

\usepackage{tikz}

\ifmodernengine\else

    \PreloadUnicodePage{0}
    \PreloadUnicodePage{3}
    \PreloadUnicodePage{32}
    \PreloadUnicodePage{33}
    \PreloadUnicodePage{34}
    \PreloadUnicodePage{37}
    \PreloadUnicodePage{39}
    \PreloadUnicodePage{468}
    \PreloadUnicodePage{471}

    \DeclareUnicodeCharacter{949}{\ensuremath\varepsilon}
    \DeclareUnicodeCharacter{952}{\ensuremath\theta}
    \DeclareUnicodeCharacter{954}{\ensuremath\kappa}
    \DeclareUnicodeCharacter{960}{\ensuremath\pi}
    \DeclareUnicodeCharacter{961}{\ensuremath\rho}
    \DeclareUnicodeCharacter{963}{\ensuremath\sigma}
    \DeclareUnicodeCharacter{966}{\ensuremath\varphi}
    \DeclareUnicodeCharacter{1013}{\ensuremath\epsilon}
    \DeclareUnicodeCharacter{977}{\ensuremath\vartheta}
    \DeclareUnicodeCharacter{1008}{\ensuremath\varkappa}
    \DeclareUnicodeCharacter{982}{\ensuremath\varpi}
    \DeclareUnicodeCharacter{1009}{\ensuremath\varrho}
    \DeclareUnicodeCharacter{962}{\ensuremath\varsigma}
    \DeclareUnicodeCharacter{981}{\ensuremath\phi}
    \new

    \DeclareUnicodeCharacter{9655}{\ensuremath\rhd}
    \DeclareUnicodeCharacter{9671}{\ensuremath\Diamond}

    \new\let\mathscr=\mathcal

\fi

\DeclareMathSymbol{.}{\mathrel}{letters}{"3A}
\DeclareMathSymbol{!}{\mathord}{operators}{"21}
\DeclareMathSymbol{?}{\mathord}{operators}{"3F}

\newcommand{\relwithsizeof}[2]{
    \mathrel{
        \text{%
            \makebox[0mm][c]{\phantom{$#1$}\makebox[0mm][c]{$#2$}}%
            \phantom{$#1$}%
        }
    }
}

\tikzset{diagram/.style={row sep=7ex,column sep=4em}}
\tikzset{object/.style={}}
\tikzset{morphism/.style={-stealth}}
\tikzset{left morphism/.style={morphism,bend left=40}}
\tikzset{right morphism/.style={morphism,bend right=40}}
\newcommand{\edges}{\AtBeginMath\scriptstyle}

\newcommand{\Ob}{\mathrm{Ob}}
\newcommand{\Mor}{\mathrm{Mor}}
\DeclareMathOperator{\Hom}{Hom}
\newcommand{\id}{\mathrm{id}}
\newcommand{\Id}{\mathrm{Id}}
\newcommand{\op}[1]{#1^{\mathrm{op}}}
\newcommand{\Set}{\mathbf{Set}}
\newcommand{\timebound}{t_{\mathrm b}}
\newcommand{\obstime}{t_{\mathrm o}}
\newcommand{\rset}{\mathbf{RSet}}
\newcommand{\ifix}{\mathbf{ifix}}

\newcommand{\event}{MSFP~’14}

\title{Categorical Semantics for Functional Reactive Programming\\
       with Temporal Recursion and Corecursion}
\author{Wolfgang Jeltsch
        \institute{TTÜ Küberneetika Instituut\\Tallinn, Estonia}
        \email{wolfgang@cs.ioc.ee}}
\def\titlerunning{Categorical Semantics for FRP with Temporal Recursion and
                  Corecursion}
\def\authorrunning{Wolfgang Jeltsch}

\begin{document}

\maketitle

\begin{abstract}

Functional reactive programming (FRP) makes it possible to express temporal
aspects of computations in a declarative way. Recently we developed two kinds of
categorical models of FRP: abstract process categories (APCs) and concrete
process categories (CPCs). Furthermore we showed that APCs generalize CPCs. In
this paper, we extend APCs with additional structure. This structure models
recursion and corecursion operators that are related to time. We show that the
resulting categorical models generalize those CPCs that impose an additional
constraint on time scales. This constraint boils down to ruling out
$ω$-supertasks, which are closely related to Zeno’s paradox of Achilles and the
tortoise.

\end{abstract}

%
%
%
%
%
%
%

\extsection{introduction}{Introduction}

Functional reactive programming (FRP) makes it possible to express temporal
aspects of computations in a declarative way. Traditional FRP is based on
behaviors and events, which denote time-varying values and values attached to
times, respectively. There is a Curry–Howard correspondence between traditional
FRP and an intuitionistic temporal logic with “always” and “eventually”
modalities~\cite{jeltsch:entcs-286,jeffrey:plpv-2012}. Thereby the type
constructor for behaviors corresponds to “always,” and the type constructor for
events corresponds to “eventually.”

Extending the temporal logic with “until” operators leads to an extended variant
of FRP. Thereby proofs of “until” propositions correspond to a new class of FRP
constructs, which we call processes. Processes in the FRP sense combine
continuous and discrete aspects and generalize behaviors and events
\cite[Section~2]{jeltsch:plpv-2013}. We give an introduction to FRP with
processes in \sectionref{functional-reactive-programming-with-processes}.

We have developed two kinds of models of FRP with processes, which can also be
used to model temporal logic with “until:”
\begin{description}

\item[Abstract process categories (APCs)~\cite{jeltsch:plpv-2014}]

are defined purely axiomatically. They are an extension of temporal
categories~\cite{jeltsch:entcs-286}, which in turn build on categorical models
of intuitionistic S4~\cite{kobayashi:tcs-175-1,bierman:sl-65-3}.

\item[Concrete process categories (CPCs)~\cite{jeltsch:plpv-2013}]

are not defined in a purely axiomatic manner, but use concrete constructions to
express time-dependence of type inhabitance and causality of FRP operations.

\end{description}
Abstract process categories generalize concrete process categories. We describe
APCs in \sectionref{abstract-process-categories} and CPCs in
\sectionref{concrete-process-categories}.

The aim of this paper is to extend APCs and CPCs in order to model recursion and
corecursion on processes. We make the following contributions:
\begin{itemize}

\item

In \sectionref{apcs-with-recursion-and-corecursion}, we develop APCs with
recursion and corecursion (ℛ-APCs). ℛ-APCs extend APCs with recursive comonad
and completely iterative monad structures that model recursion and corecursion
on processes. The extensions to APCs arise naturally as extensions of ideal
comonad and ideal monad structures that APCs already contain.

\item

In \sectionref{cpcs-with-recursion-and-corecursion}, we develop CPCs with
recursion and corecursion (ℛ-CPCs). ℛ-CPCs differ from CPCs in that they impose
an additional constraint on time scales. This constraint boils down to ruling
out $ω$-supertasks~\cite{laraudogoitia:supertasks}. We show that ℛ-APCs
generalize ℛ-CPCs.

\end{itemize}

We discuss related work in \sectionref{related-work} and give conclusions and an
outlook on further work in \sectionref{conclusions-and-further-work}.

\extsection{functional-reactive-programming-with-processes}
           {Functional Reactive Programming with Processes}

The FRP dialect we consider here is based on a linear notion of time. However
the time scale does not need to be discrete; so the time domain can be any
totally ordered set. Type inhabitance is time-dependent, which means that every
FRP type corresponds to a time-indexed family of sets of inhabitants.

Besides the ordinary type constructors $1$, $0$, $×$, $+$, and~$→$, our FRP
dialect has two binary type constructors $▷″₀$ and~$▷″₁$, which we call the
strong and the weak basic process type constructor. Both have higher precedence
than the other binary type constructors.

A value that inhabits a type $τ₁ ▷″₀ τ₂$ at a time~$t$ corresponds to a tuple
with the following elements:
\begin{itemize}

\item

a time~$t′ > t$

\item

a function~$h$ that maps each time $t″$ with $t < t″ < t′$ to a value that
inhabits $τ₁$ at~$t″$

\item

a value~$y$ that inhabits $τ₂$ at~$t′$

\end{itemize}
The function~$h$ denotes a time-varying value that exists between times $t$
and~$t′$. We call this time-varying value the continuous part of the process.
The pair $(t′, y)$ denotes an event that occurs at time~$t′$ and carries the
value~$y$. We call this event the terminal event of the process.

A value that inhabits a type $τ₁ ▷″₁ τ₂$ at a time~$t$ is either a process that
inhabits $τ₁ ▷″₀ τ₂$ at~$t$ or a time-varying value of type~$τ₁$ that starts
immediately after~$t$ and persists forever. We regard constructs of the latter
kind as special processes that never terminate and thus have an infinite
continuous part and no terminal event.

An inhabitant of a type $τ₁ ▷″₀ τ₂$ or $τ₁ ▷″₁ τ₂$ has a continuous part that
assigns values only to future times. We define type constructors $▷′₀$ and~$▷′₁$
for processes that start at the present time:
\begin{align}
τ₁ ▷′₀ τ₂ & = τ₁ × τ₁ ▷″₀ τ₂ & τ₁ ▷′₁ τ₂ & = τ₁ × τ₁ ▷″₁ τ₂
\end{align}
A pair $(x, p)$ represents a process that starts with~$x$ as the initial value
of its continuous part, and continues like the process~$p$.

An inhabitant of a type $τ₁ ▷′₀ τ₂$ or $τ₁ ▷′₁ τ₂$ can only terminate in the
future. We define type constructors $▷₀$ and~$▷₁$ for processes that may
terminate at the present time:
\begin{align}
τ₁ ▷₀ τ₂ & = τ₂ + τ₁ ▷′₀ τ₂ & τ₁ ▷₁ τ₂ & = τ₂ + τ₁ ▷′₁ τ₂
\end{align}
A value $ι₁(y)$ represents a process whose terminal event occurs immediately and
carries the value~$y$, while a value $ι₂(p)$ represents the process~$p$, which
does not terminate immediately.

From the process type constructors, we derive the type constructors $□′$, $□$,
$◇′$, and~$◇$ as follows:
\begin{align}
□′τ & = τ ▷″₁ 0 & □τ & = τ ▷′₁ 0 \\
◇′τ & = 1 ▷′₀ τ & ◇τ & = 1 ▷₀ τ
\end{align}
Inhabitants of types $□′τ$ and $□τ$ are non-terminating time-varying values and
are called behaviors. Inhabitants of types $◇′τ$ and $◇τ$ are events.

The FRP dialect described above corresponds to an intuitionistic temporal logic
via a Curry–Howard isomorphism. Thereby time-dependent type inhabitance is
related to time-dependent trueness of temporal propositions. The type
constructors $▷₀$ and~$▷₁$ correspond to the strong and weak “until” operators
$𝒰$ and~$𝒲$ from linear-time temporal logic (LTL), and the type constructors $□$
and~$◇$ correspond to the “always” and “eventually” modalities.

\extsection{abstract-process-categories}{Abstract Process Categories}

Abstract process categories (APCs) are axiomatically defined categorical models
of FRP with processes, which we developed recently~\cite{jeltsch:plpv-2014}.
They are an extension of temporal categories, which are models of FRP with
behaviors and events, but without processes~\cite{jeltsch:entcs-286}. In this
section, we give an introduction to APCs.

\extsubsection{the-basics}{The Basics}

An APC is a category~$𝒞$ with some additional structure. FRP types are modeled
by objects of~$𝒞$. If objects $A$ and~$B$ model FRP types $τ₁$ and~$τ₂$, the
morphisms from~$A$ to~$B$ model operations that turn every value that
inhabits~$τ₁$ at some time into a value that inhabits~$τ₂$ at the same time.

Since FRP has the usual type constructors for finite products, finite sums, and
function spaces, we require~$𝒞$ to be a cartesian closed category (CCC) with
finite coproducts.

\extsubsection{temporal-functors}{Temporal Functors}

Let $𝟐$ denote the interval category, that is, the category with exactly two
objects $0$ and~$1$ and a single non-identity morphism $w : 0 → 1$. Let
furthermore $𝒬$ be the category $𝟐 × 𝒞 × 𝒞$. We require the existence of a
functor $▷″ : 𝒬 → 𝒞$, which we call the basic temporal functor. We use the
notation $A ▷″_W B$ for $▷″(W, A, B)$. We model the process type constructors
$▷″₀$ and~$▷″₁$ by the partial functor applications $▷″₀$ and~$▷″₁$, which are
themselves functors from $𝒞 × 𝒞$ to~$𝒞$.

Every inhabitant of a type $τ₁ ▷″₀ τ₂$ corresponds to an inhabitant of $τ₁ ▷″₁
τ₂$. So there should be an operation that performs a type conversion from $▷″₀$
to~$▷″₁$. We use the natural transformation $▷″_w : ▷″₀ → ▷″₁$ to model this
operation.

From the functor~$▷″$, we derive functors $▷′, ▷ : 𝒬 → 𝒞$ that model the type
constructors $▷′₀$, $▷′₁$, $▷₀$, and~$▷₁$ as well as functors $□′, □, ◇′, ◇ : 𝒞
→ 𝒞$ that model the type constructors for behaviors and events:
\begin{align}
\label{equation:derived-process-functors}
A ▷′_W B & = A × A ▷″_W B & A ▷_W B  & = B + A ▷′_W B \\
□′A & = A ▷″₁ 0 & □A & = A ▷′₁ 0                      \\
\label{equation:event-functors}
◇′B & = 1 ▷′₀ B & ◇B & = 1 ▷₀ B
\end{align}
We call the functors introduced in \eqref{equation:derived-process-functors}
through~\eqref{equation:event-functors} derived temporal functors.

APCs do not necessarily use the category~$𝟐$ to model how weak a process type
constructor is. Any category~$𝒲$ with finite products is allowed instead. In
particular, we can use any category that corresponds to a partially ordered set
with finite meets. This makes it possible to model more advanced termination
properties. An example of such a property is termination with an upper bound on
the termination time. We discuss this example in
\subsectionref{the-basic-temporal-functor}.

\extsubsection{process-expansion-and-joining}{Process Expansion and Joining}

An APC contains a natural transformation~$θ″$ with
\begin{equation*}
θ″_{W, A, B} : A ▷″_W B → \left(A ▷′_W B\right) ▷″_W B\enspace.
\end{equation*}
This natural transformation models an FRP operation that turns a process~$p$
into a process~$p′$ such that the following holds:
\begin{itemize}

\item

The process~$p′$ terminates if and only if~$p$ terminates.

\item

The terminal event of~$p′$, if any, is the terminal event of~$p$.

\item

The value of the continuous part of~$p′$ at a time~$t$ is the suffix of~$p$ that
starts at~$t$.

\end{itemize}
We call this FRP operation process expansion. We derive variants of~$θ″$ that
work with $▷′$ and~$▷$ instead of~$▷″$:
\begin{align}
θ′_{W, A, B} & \relwithsizeof=: A ▷′_W B → \left(A ▷′_W B\right) ▷′_W B                    \notag\\
θ′_{W, A, B} & =                \left\langle\id_{A ▷′_W B}, θ″_{W, A, B} ∘ π₂\right\rangle \\[\topsep]
θ_{W, A, B}  & \relwithsizeof=: A ▷_W B → \left(A ▷′_W B\right) ▷_W B                      \notag\\
\label{equation:no-prime-theta}
θ_{W, A, B}  & =                \id_B + θ′_{W, A, B}
\end{align}

Besides the natural transformation~$θ″$, an APC also contains a natural
transformation~$ϑ″$ with
\begin{equation*}
ϑ″_{W, A, B} : A ▷″_W (A ▷_W B) → A ▷″_W B\enspace.
\end{equation*}
This natural transformation models an FRP operation that turns a process~$p$
into a process~$p′$ such that the following holds:
\begin{itemize}

\item

If $p$ terminates, and its terminal event carries a process~$p^*$, then $p′$ is
the result of concatenating the continuous part of~$p$ and the process~$p^*$.

\item

If $p$ does not terminate, then $p′$ does not terminate, and the continuous part
of~$p′$ is the continuous part of~$p$.

\end{itemize}
We call this FRP operation process joining. We derive variants of~$ϑ″$ that work
with $▷′$ and~$▷$ instead of~$▷″$:
\begin{align}
ϑ′_{W, A, B} & \relwithsizeof=: A ▷′_W (A ▷_W B) → A ▷′_W B                   \notag\\
\label{equation:single-prime-vartheta}
ϑ′_{W, A, B} & =                \id_A × ϑ″_{W, A, B}                          \\[\topsep]
ϑ_{W, A, B}  & \relwithsizeof=: A ▷_W (A ▷_W B) → A ▷_W B                     \notag\\
\label{equation:no-prime-vartheta}
ϑ_{W, A, B}  & =                \left[\id_{A ▷_W B}, ι₂ ∘ ϑ′_{W, A, B}\right]
\end{align}

The natural transformations $θ″$ and~$ϑ″$ have to fulfill certain coherence
conditions. Regarding $θ″$, we require that for all $W ∈ \Ob(𝒲)$ and $B ∈
\Ob(𝒞)$, $\left({-} ▷″_W B, θ″_{W, {-}, B}\right)$ is an ideal comonad.

\begin{extdefinition}{ideal-comonad}[Ideal comonad]
Let $𝒞$ be a category with binary products, let $U′$ be an endofunctor on~$𝒞$,
and let $δ′ : U′ → U′(\Id × U′)$ be a natural transformation. We define $U$,
$ε$, and~$δ$ as follows:
\begin{align}
U & \relwithsizeof=: 𝒞 → 𝒞            &
ε & \relwithsizeof=: U → \Id          &
δ & \relwithsizeof=: U → UU           \notag\\
U & =                \Id × U′         &
ε & =                π₁               &
δ & =                ⟨\id_U, δ′ ∘ π₂⟩
\end{align}
The pair $(U′, δ′)$ is an ideal comonad on~$𝒞$ if and only if the diagram in
\figureref{coherence-of-an-ideal-comonad} commutes.
\end{extdefinition}
\begin{extfigure}{coherence-of-an-ideal-comonad}{Coherence of an ideal comonad}

\begin{tikzpicture}[diagram]

\matrix{
\node [object] (epsilon end)           {$U′$};   &
\node [object] (inside source)         {$U′U$};  &
\node [object] (delta end)             {$U′UU$}; \\
                                                 &
\node [object] (start)                 {$U′$};   &
\node [object] (outside delta source)  {$U′U$};  \\
};

\edges

\path (start)                edge [morphism] node [auto=right] {$δ′$}       (inside source)
      (inside source)        edge [morphism] node [auto=right] {$U′ε$}      (epsilon end)
      (start)                edge [morphism] node [auto=left]  {$\id_{U′}$} (epsilon end)
      (inside source)        edge [morphism] node [auto=left]  {$U′δ$}      (delta end)
      (start)                edge [morphism] node [auto=right] {$δ′$}       (outside delta source)
      (outside delta source) edge [morphism] node [auto=right] {$δ′U$}      (delta end);

\end{tikzpicture}

\end{extfigure}

Regarding $ϑ″$, we require that the diagram in
\figureref{coherence-of-process-joining} commutes. This coherence condition
implies that every pair $\left(A ▷′_W {-}, ϑ′_{W, A, {-}}\right)$ is an ideal
monad, that is, an ideal comonad on~$\op{𝒞}$. Note that we obtain a diagram for
the coherence condition of such an ideal monad by taking the diagram in
\figureref{coherence-of-process-joining} and replacing $▷″$ and~$ϑ″$ with $▷′$
and~$ϑ′$.
\begin{extfigure}{coherence-of-process-joining}{Coherence of process joining}

\begin{tikzpicture}[diagram,column sep=6.3em]

\matrix{
\node [object] (eta start)              {$A ▷″_W B$};                 &
\node [object] (inside destination)     {$A ▷″_W (A ▷_W B)$};         &
\node [object] (mu start)               {$A ▷″_W (A ▷_W (A ▷_W B))$}; \\
                                                                      &
\node [object] (end)                    {$A ▷″_W B$};                 &
\node [object] (outside mu destination) {$A ▷″_W (A ▷_W B)$};         \\
};

\edges

\path (eta start)              edge [morphism] node [auto=left]  {$A ▷″_W ι₁$}          (inside destination)
      (inside destination)     edge [morphism] node [auto=left]  {$ϑ″_{W, A, B}$}       (end)
      (eta start)              edge [morphism] node [auto=right] {$\id_{A ▷″_W B}$}     (end)
      (mu start)               edge [morphism] node [auto=right] {$A ▷″_W ϑ_{W, A, B}$} (inside destination)
      (mu start)               edge [morphism] node [auto=left]  {$ϑ″_{W, A, A ▷_W B}$} (outside mu destination)
      (outside mu destination) edge [morphism] node [auto=left]  {$ϑ″_{W, A, B}$}       (end);

\end{tikzpicture}

\end{extfigure}

Finally we require that the diagram in
\figureref{coherence-between-process-expansion-and-joining} commutes. This
ensures that $θ″$ and~$ϑ″$ interact properly.
\begin{extfigure}{coherence-between-process-expansion-and-joining}
                 {Coherence between process expansion and joining}

\begin{tikzpicture}[diagram,column sep=7em]

\matrix{
\node [object] (nested on the right)        {$A ▷″_W (A ▷_W B)$};                                                    &
\node [object] (deeper nested on the left)  {$\left(A ▷′_W (A ▷_W B)\right) ▷″_W (A ▷_W B)$};                        \\
\node [object] (not nested)                 {$A ▷″_W B$};                                                            &
                                                                                                                     \\
\node [object] (nested on the left)         {$\left(A ▷′_W B\right) ▷″_W B$};                                        &
\node [object] (deeper nested on the right) {$\left(A ▷′_W B\right) ▷″_W \left(\left(A ▷′_W B\right) ▷_W B\right)$}; \\
};

\edges

\path (nested on the right)        edge [morphism] node [auto=left]  {$θ″_{W, A, A ▷_W B}$}            (deeper nested on the left)
      (deeper nested on the left)  edge [morphism] node [auto=left]  {$ϑ′_{W, A, B} ▷″_W θ_{W, A, B}$} (deeper nested on the right)
      (deeper nested on the right) edge [morphism] node [auto=left]  {$ϑ″_{W, A ▷′_W B, B}$}           (nested on the left)
      (nested on the right)        edge [morphism] node [auto=right] {$ϑ″_{W, A, B}$}                  (not nested)
      (not nested)                 edge [morphism] node [auto=right] {$θ″_{W, A, B}$}                  (nested on the left);

\end{tikzpicture}

\end{extfigure}

\extsubsection{process-merging-and-the-canonical-nonterminating-process}
              {Process Merging and the Canonical Nonterminating Process}

There are additional constraints on APCs, which ensure that an APC also models
two other operations: process merging and the construction of the sole value
of~$1 ▷″₁ 0$, which is called the canonical nonterminating process. We do not
discuss these additional constraints here, because they are not fundamentally
related to our extension of APCs. We just mention them in the definition of APCs
in the next subsection.

\extsubsection{abstract-process-categories-summary}{Summary}

Let us now state the complete definition of abstract process categories.

\begin{extdefinition}{abstract-process-category}[Abstract process category]
Let $𝒲$ be a category with finite products, and let $𝒞$ be a CCC with finite
coproducts. Furthermore let $▷″$ be a functor from~$𝒲 × 𝒞 × 𝒞$ to~$𝒞$, let $▷′$
and~$▷$ be defined as in \eqref{equation:derived-process-functors}, and let $θ″$
and~$ϑ″$ be natural transformations with the following typings:
\begin{align*}
θ″_{W, A, B} & : A ▷″_W B → \left(A ▷′_W B\right) ▷″_W B \\
ϑ″_{W, A, B} & : A ▷″_W (A ▷_W B) → A ▷″_W B
\end{align*}
The tuple $(𝒲, 𝒞, ▷″, θ″, ϑ″)$ is an abstract process category (APC) if and only
if the following propositions hold:
\begin{itemize}

\item

For all $W ∈ \Ob(𝒲)$ and $B ∈ \Ob(𝒞)$, $\left({-} ▷″_W B, θ″_{W, {-}, B}\right)$
is an ideal comonad.

\item

For all $W ∈ \Ob(𝒲)$ and $A, B ∈ \Ob(𝒞)$, the diagrams in Figures
\ref{figure:coherence-of-process-joining}
and~\ref{figure:coherence-between-process-expansion-and-joining} commute where
$θ$, $ϑ′$, and~$ϑ$ are defined as in \eqref{equation:no-prime-theta},
\eqref{equation:single-prime-vartheta}, and~\eqref{equation:no-prime-vartheta}.

\item

The natural transformation $\left⟨χ″₁, χ″₂\right⟩$ with
\begin{align}
χ″_i & \relwithsizeof=: (A₁ × A₂) ▷″_{W₁ × W₂} \left(B₁ × B₂ + B₁ × A₂ ▷′_{W₂} B₂ + A₁ ▷′_{W₁} B₁ × B₂\right) \notag\\
     & \qquad→          A_i ▷″_{W_i} B_i                                                                      \notag\\
χ″_i & =                ϑ″_{W_i, A_i, B_i} ∘ π_i ▷″_{π_i} [ι₁ ∘ π_i, ι_i ∘ π_i, ι_{3 - i} ∘ π_i]
\end{align}
is an isomorphism.

\item

The morphism $!_{1 ▷″₁ 0}$ is an isomorphism.

\end{itemize}
\end{extdefinition}

The third and fourth of the above propositions refer to process merging and the
canonical nonterminating process mentioned in
\subsectionref{process-merging-and-the-canonical-nonterminating-process}.

\extsection{apcs-with-recursion-and-corecursion}
           {APCs with Recursion and Corecursion}

In \subsectionref{process-expansion-and-joining}, we saw that every APC
comprises ideal comonads and ideal monads based on temporal functors. There
exists a special kind of ideal comonads, called recursive comonads, which
captures a form of comonadic recursion. Likewise there exists a special kind of
ideal monads, called completely iterative monads, which captures a form of
monadic corecursion. In this section, we introduce APCs with recursion and
corecursion (ℛ-APCs), which comprise recursive comonads and completely iterative
monads that model recursion and corecursion on processes.

\extsubsection{corecursion-on-processes}{Corecursion on Processes}

If $(𝒲, 𝒞, ▷″, θ″, ϑ″)$ is an APC, then every pair $\left(A ▷′_W {-}, ϑ′_{W, A,
{-}}\right)$ is an ideal monad. We extend the structure of APCs by requiring
that every such pair is a completely iterative monad. Completely iterative
monads are defined, for example, by Milius
\cite[Definition~5.5]{milius:ic-196-1}. We use a definition that is different
from the one by Milius, but nevertheless equivalent to it. We compare both
definitions in
\appendixref{changes-in-the-completely-iterative-monad-definition}, where we
also list the advantages that our definition has in our opinion.

\begin{extdefinition}{completely-iterative-monad}[Completely iterative monad]
Let $𝒞$ be a category with binary coproducts. A pair $(T′, μ′)$ is a completely
iterative monad on~$𝒞$ if and only if it is an ideal monad on~$𝒞$, and for every
morphism $f : C → T′(B + C)$, there exists a unique morphism $f^∞ : C → T′B$ for
which the diagram in
\figureref{condition-for-a-morphism-f-infinity-of-a-completely-iterative-monad}
commutes.
\end{extdefinition}
\begin{extfigure}{condition-for-a-morphism-f-infinity-of-a-completely-iterative-monad}
                 {Condition for a morphism~$f^∞$ of a completely iterative monad}

\begin{tikzpicture}[diagram,column sep=7em]

\matrix{
\node [object] (start)            {$C$};           &
\node [object] (end)              {$T′B$};         \\
\node [object] (after expansion)  {$T′(B + C)$};   &
\node [object] (after inner call) {$T′(B + T′B)$}; \\
};

\edges

\path (start)            edge [morphism] node [auto=left]  {$f^∞$}             (end)
      (start)            edge [morphism] node [auto=right] {$f$}               (after expansion)
      (after expansion)  edge [morphism] node [auto=right] {$T′(\id_B + f^∞)$} (after inner call)
      (after inner call) edge [morphism] node [auto=right] {$μ′_B$}            (end);

\end{tikzpicture}

\end{extfigure}

By requiring that every pair $\left(A ▷′_W {-}, ϑ′_{W, A, {-}}\right)$ is a
completely iterative monad, we ensure that for every morphism $f : C → A ▷′_W (B
+ C)$, there is a corresponding morphism $f^∞ : C → A ▷′_W B$. Let $u$ and~$u^∞$
be the FRP operations that $f$ and~$f^∞$ model, and let $z$ be a value from the
domain of $u$ and~$u^∞$. The diagram in
\figureref{condition-for-a-morphism-f-infinity-of-a-completely-iterative-monad}
tells us how the process $u^∞(z)$ is defined:
\begin{itemize}

\item

If $u(z)$ terminates, and its terminal event carries a value $ι₁(y)$, then
$u^∞(z)$ terminates at the same time as $u(z)$, the continuous part of $u^∞(z)$
is the continuous part of $u(z)$, and the terminal event of $u^∞(z)$ carries the
value~$y$.

\item

If $u(z)$ terminates, and its terminal event carries a value $ι₂(z′)$, then
$u^∞(z)$ is the result of concatenating the continuous part of $u(z)$ and the
process~$u^∞(z′)$.

\item

If $u(z)$ does not terminate, then $u^∞(z)$ does not terminate, and the
continuous part of $u^∞(z)$ is the continuous part of $u(z)$.

\end{itemize}
Note that in the second case, $u^∞(z)$ is defined in terms of $u^∞(z′)$, but
only the suffix of $u^∞(z)$ that starts at the termination time of $u(z)$
actually depends on $u^∞(z′)$. The operator~${-}^∞$ models corecursion on
processes.

We derive a variant of~${-}^∞$ that works with~$▷$ instead of~$▷′$. We use the
notation ${-}^∞$ also for this variant. This overloading of notation is
possible, because we can always deduce from the types which variant of~${-}^∞$
is meant. The $▷$-variant turns every morphism $f : C → B + A ▷′_W C$ into a
morphism~$f^∞$ that is defined as follows:
\begin{align}
f^∞ & \relwithsizeof=: C → A ▷_W B                                                \notag\\
f^∞ & =                \left(\id_B + \left(\id_A ▷′_{\id_W} f\right)^∞\right) ∘ f
\end{align}
Note that the ${-}^∞$ on the right-hand side refers to the original
$▷′$-variant. There is also a variant of~${-}^∞$ that works with~$▷″$. It turns
every morphism $f : C → A ▷″_W (B + A × C)$ into a morphism~$f^∞$ that is
defined as follows:
\begin{align}
f^∞ & \relwithsizeof=: C → A ▷″_W B                                                             \notag\\
f^∞ & =                ϑ″_{W, A, B} ∘ \left(\id_A ▷″_{\id_W} (\id_B + (\id_A × f)^∞)\right) ∘ f
\end{align}
Again the ${-}^∞$ on the right-hand side refers to the original $▷′$-variant.

\extsubsection{recursion-on-processes}{Recursion on Processes}

We saw in \subsectionref{process-expansion-and-joining} that every pair
$\left({-} ▷″_W B, θ″_{W, {-}, B}\right)$ is an ideal comonad. We extend the
structure of APCs by requiring that every such pair is a recursive comonad. The
notion of recursive comonad is dual to the notion of completely iterative monad.
So a recursive comonad on~$𝒞$ is just a completely iterative monad on~$\op{𝒞}$.
We give an explicit definition nevertheless.

\begin{extdefinition}{recursive-comonad}[Recursive comonad]
Let $𝒞$ be a category with binary products. A pair $(U′, δ′)$ is a recursive
comonad on~$𝒞$ if and only if it is an ideal comonad on~$𝒞$, and for every
morphism $f : U′(A × C) → C$, there exists a unique morphism $f^* : U′A → C$ for
which the diagram in
\figureref{condition-for-a-morphism-f-star-of-a-recursive-comonad} commutes.
\end{extdefinition}
\begin{extfigure}{condition-for-a-morphism-f-star-of-a-recursive-comonad}
                 {Condition for a morphism~$f^*$ of a recursive comonad}

\begin{tikzpicture}[diagram,column sep=7em]

\matrix{
\node [object] (end)               {$C$};           &
\node [object] (start)             {$U′A$};         \\
\node [object] (before collapsing) {$U′(A × C)$};   &
\node [object] (before inner call) {$U′(A × U′A)$}; \\
};

\edges

\path (start)             edge [morphism] node [auto=right] {$f^*$}                        (end)
      (before collapsing) edge [morphism] node [auto=left]  {$f$}                          (end)
      (before inner call) edge [morphism] node [auto=left]  {$U′\left(\id_A × f^*\right)$} (before collapsing)
      (start)             edge [morphism] node [auto=left]  {$δ′_A$}                       (before inner call);

\end{tikzpicture}

\end{extfigure}

By requiring that every pair $\left({-} ▷″_W B, θ″_{W, {-}, B}\right)$ is a
recursive comonad, we ensure that for every morphism $f : (A × C) ▷″_W B → C$,
there is a corresponding morphism $f^* : A ▷″_W B → C$. Let $u$ and~$u^*$ be the
FRP operations that $f$ and~$f^*$ model, and let $p$ be a process from the
domain of~$u^*$. The diagram in
\figureref{condition-for-a-morphism-f-star-of-a-recursive-comonad} tells us that
the value $u^*(p)$ is $u\left(p^†\right)$ where $p^†$ is defined by the
following statements:
\begin{itemize}

\item

The process~$p^†$ terminates if and only if~$p$ terminates.

\item

The terminal event of~$p^†$, if any, is the terminal event of~$p$.

\item

The value of the continuous part of~$p^†$ at a time~$t$ is $(x, u^*(p′))$ where
$x$ is the value of the continuous part of~$p$ at~$t$, and $p′$ is the suffix
of~$p$ that follows~$t$.

\end{itemize}
Note that $u^*(p)$ is defined in terms of $u^*(p′)$, but $p′$ is a proper suffix
of~$p$. The operator~${-}^*$ models recursion on processes.

We derive a variant of~${-}^*$ that works with~$▷′$ instead of~$▷″$. We use the
notation ${-}^*$ also for this variant. The $▷′$-variant turns every morphism $f
: A × C ▷″_W B → C$ into a morphism~$f^*$ that is defined as follows:
\begin{align}
f^* & \relwithsizeof=: A ▷′_W B → C                                               \notag\\
f^* & =                f ∘ \left(\id_A × \left(f ▷″_{\id_W} \id_B\right)^*\right)
\end{align}
Note that this construction is dual to the construction of the $▷$-variant
of~${-}^∞$.

\extsubsection{apcs-with-recursion-and-corecursion-summary}{Summary}

We formulate the definition of ℛ-APCs based on the above explanations.

\begin{extdefinition}{apc-with-recursion-and-corecursion}
                     [APC with recursion and corecursion]
An APC with recursion and corecursion (ℛ-APC) is an APC $(𝒲, 𝒞, ▷″, θ″, ϑ″)$ for
which the following holds:
\begin{itemize}

\item

For all $W ∈ \Ob(𝒲)$ and $B ∈ \Ob(𝒞)$, $\left({-} ▷″_W B, θ″_{W, {-}, B}\right)$
is a recursive comonad.

\item

For all $W ∈ \Ob(𝒲)$ and $A ∈ \Ob(𝒞)$, $\left(A ▷′_W {-}, ϑ′_{W, A, {-}}\right)$
is a completely iterative monad.

\end{itemize}
\end{extdefinition}

\extsection{concrete-process-categories}{Concrete Process Categories}

Concrete process categories (CPCs) are models of FRP with processes that use
concrete categorical constructions to express time-dependence of type
inhabitance and causality of FRP operations. They are described in detail in an
earlier publication of ours \cite[Section~3]{jeltsch:plpv-2013}. Here we only
give a short introduction to CPCs.

\extsubsection{core-structure}{Core Structure}

Let $(T, ≤)$ be a totally ordered set, and let $ℬ$ be a CCC with finite
coproducts. We use $(T, ≤)$ to model the time scale, and $ℬ$ to model ordinary
types and functions. Based on $(T, ≤)$, we construct a category~$ℐ$, which we
call the temporal index category of $(T, ≤)$.

\begin{extdefinition}{temporal-index-category}[Temporal index category]
The temporal index category of a totally ordered set $(T, ≤)$ is the
category~$ℐ$ for which the following holds:
\begin{align}
\Ob(ℐ)                                                  &
= \set{(t, \obstime) ∈ T × T}{t ≤ \obstime}             \\
\hom_ℐ\bigl((t′,\obstime′),(t, \obstime)\bigr)          &
= \begin{cases}
  \left\{\left(t, \obstime, \obstime′\right)\right\} &
  \text{if $t = t′$ and $\obstime ≤ \obstime′$}      \\
  ∅                                                  &
  \text{otherwise}
  \end{cases}
\end{align}
\end{extdefinition}

We model FRP types and FRP operations by the functor category~$ℬ^ℐ$. Let $A$ be
an object of~$ℬ^ℐ$ that models an FRP type~$τ$. For every object $(t, \obstime)$
of~$ℐ$, the object $A(t, \obstime)$ of~$ℬ$ deals with the FRP values that
inhabit~$τ$ at~$t$; it describes the type whose inhabitants give the information
we have about these FRP values at~$\obstime$. We call $\obstime$ the observation
time. For every morphism $\left(t, \obstime, \obstime′\right)$ of~$ℐ$, the
morphism $A\left(t, \obstime, \obstime′\right)$ models the function that turns
information we have at~$\obstime′$ into information we have at~$\obstime$ by
forgetting all information acquired after~$\obstime$

The definition of CPCs requires the category~$ℬ$ to have some additional
properties, which are necessary for modeling function types and process types in
FRP.

\begin{extdefinition}{concrete-process-category}[Concrete process category]
Let $(T, ≤)$ be a totally ordered set, let $ℐ$ be its temporal index category,
and let $ℬ$ be a CCC with finite coproducts that has all products and coproducts
of families indexed by intervals $(t, t′) ⊆ T$ and all ends of the form $∫_{t″ ∈
[t, t′]} B(t, t″)^{A\left(t, t″\right)}$ where $A, B ∈ ℬ^ℐ$. Then the tuple $(T,
≤, ℬ)$ is a concrete process category (CPC). The actual category that $(T, ≤,
ℬ)$ denotes is the functor category~$ℬ^ℐ$.
\end{extdefinition}

\extsubsection{the-basic-temporal-functor}{The Basic Temporal Functor}

The basic temporal functor of a CPC \cite[Subsection~4.3]{jeltsch:plpv-2014}
models the weak basic process type constructor as well as process type
constructors that put upper bounds on termination times.

We define the totally ordered set $(T_∞, ≤_∞)$ such that $T_∞ = T ∪ \{∞\}$ and
\begin{equation*}
t₁ ≤_∞ t₂ ⇔ t₁ ≤ t₂ ∨ t₂ = ∞\enspace.
\end{equation*}
Let $𝒲$ be the category of $(T_∞, ≤_∞)$. We use $𝒲$ to model constraints
regarding termination. Thereby any object $\timebound ∈ T$ stands for
termination at or before the time~$\timebound$, and the object~$∞$ stands for
the absence of any termination guarantees. We define the basic temporal functor
as follows.

\begin{extdefinition}{basic-temporal-functor-of-a-cpc}
                     [Basic temporal functor of a CPC]
Let $(T, ≤, ℬ)$ be a CPC, and let $𝒲$ be defined as above. The basic temporal
functor of~$(T, ≤, ℬ)$ is the functor $▷″ : 𝒲 × ℬ^ℐ × ℬ^ℐ → ℬ^ℐ$ with
\begin{equation}
\label{equation:process-information-objects}
\left(A ▷″_{\timebound} B \right)(t, \obstime)
    = \begin{cases}
      0                                                     & \text{if $\timebound < t$}            \\
      S_{\timebound}                                        & \text{if $t ≤ \timebound ≤ \obstime$} \\
      S_{\obstime} + ∏_{t′ ∈ (t, \obstime]} A(t′, \obstime) & \text{if $\obstime <_∞ \timebound$}
      \end{cases}
\end{equation}
where for every~$t^* ∈ T$, $S_{t^*}$ is defined as follows:
\begin{equation}
\label{equation:process-information-objects-helper}
S_{t^*} = ∐_{t′ ∈ (t, t^*]} \left(\left(∏_{t″ ∈ (t, t′)} A(t″, \obstime)\right) × B(t′, \obstime)\right)
\end{equation}
\end{extdefinition}

The above definition is actually incomplete. A complete definition would also
define the following:
\begin{itemize}

\item

morphisms $\left(A ▷″_W B\right)i$ where $W ∈ \Ob(𝒲)$, $A, B ∈
\Ob\left(ℬ^ℐ\right)$, and $i ∈ \Mor(ℐ)$

\item

morphisms $\left(f ▷″_{\id_W} g\right)_I$ where $W ∈ \Ob(𝒲)$, $f, g ∈
\Mor\left(ℬ^ℐ\right)$, and $I ∈ \Ob(ℐ)$

\item

morphisms $\left(\id_A ▷″_w \id_B\right)_I$ where $w ∈ \Mor(𝒲)$, $A, B ∈
\Ob\left(ℬ^ℐ\right)$, and $I ∈ \Ob(ℐ)$

\end{itemize}
However defining these things is a mostly mechanical task. In particular, the
definition of morphisms $\left(f ▷″_{\id_W} g\right)_I$ can be derived from
\eqref{equation:process-information-objects}
and~\eqref{equation:process-information-objects-helper} by replacing $0$ with
$\id_0$ and object expressions of the form $C(t, \obstime)$ with morphism
expressions of the form $h_{(t, \obstime)}$.

\extsubsection{relationship-to-abstract-process-categories}
              {Relationship to Abstract Process Categories}

APCs generalize CPCs. This is expressed by the following theorem.

\begin{exttheorem}{apc-from-cpc}
Let $(T, ≤, ℬ)$ be a CPC, let $ℐ$ be the temporal index category of $(T, ≤)$,
let $▷″$ be the basic temporal functor of $(T, ≤, ℬ)$, and let $𝒲$ be the
category of the totally ordered set $(T_∞, ≤_∞)$ that is defined as described in
\subsectionref{the-basic-temporal-functor}. Then there exist natural
transformations $θ″$ and~$ϑ″$ such that $\left(𝒲, ℬ^ℐ, ▷″, θ″, ϑ″\right)$ is an
APC.
\end{exttheorem}

Please see our earlier work \cite[proof of Theorem~10]{jeltsch:plpv-2014} for a
proof of this theorem.

\extsection{cpcs-with-recursion-and-corecursion}
           {CPCs with Recursion and Corecursion}

According to \theoremref{apc-from-cpc}, every CPC gives rise to an APC. However
there is no guarantee that this APC is also an ℛ-APC. In this section, we
develop CPCs with recursion and corecursion (ℛ-CPCs). ℛ-CPCs are a special kind
of CPCs whose corresponding APCs are ℛ-APCs.

\extsubsection{enabling-corecursion-on-processes}
              {Enabling Corecursion on Processes}

Let $(T, ≤, ℬ)$ be a CPC. Let furthermore $\left(𝒲, ℬ^ℐ, ▷″, θ″, ϑ″\right)$ be
the APC it induces according to \theoremref{apc-from-cpc}. This APC models
corecursion on processes if and only if for every morphism $f : C → A
▷′_{\timebound} (B + C)$, there exists a unique morphism~$f^∞$ for which the
following holds:
\begin{align}
f^∞ & \relwithsizeof=: C → A ▷′_{\timebound} B                                  \notag\\
\label{equation:f-infinity-for-apc}
f^∞ & =                ϑ′_{\timebound, A, B}                                  ∘
                       \left(\id_A ▷′_{\id_{\timebound}} (\id_B + f^∞)\right) ∘
                       f
\end{align}
From the definition of~$▷′$ in \eqref{equation:derived-process-functors} and the
definition of the basic temporal functor in
\subsectionref{the-basic-temporal-functor}, it follows that
\eqref{equation:f-infinity-for-apc} is true if and only if for all $t, \obstime
∈ T$ with $t ≤ \obstime$, we have
\begin{equation}
\label{equation:f-infinity-element-definition}
f^∞_{(t, \obstime)} = \left(ϑ′_{\timebound, A, B}\right)_{(t, \obstime)} ∘
                      (\id_A × h)                                        ∘
                      f_{(t, \obstime)}
\end{equation}
where $h$ is defined as follows:
\begin{equation}
\label{equation:f-infinity-element-definition-helper}
h = \begin{cases}
    \id_0                                                                                                &
        \text {if $\timebound < t$}                                                                      \\
    ∐_{t′ ∈ (t, \timebound]} \left(\id × \left(\id + f^∞_{(t′, \obstime)}\right)\right)                  &
        \text {if $t ≤ \timebound ≤ \obstime$}                                                           \\[\medskipamount]
    \left(∐_{t′ ∈ (t, \obstime]} \left(\id × \left(\id + f^∞_{(t′, \obstime)}\right)\right)\right) + \id &
        \text {if $\obstime <_∞ \timebound$}
    \end{cases}
\end{equation}
Equations \eqref{equation:f-infinity-element-definition}
and~\eqref{equation:f-infinity-element-definition-helper} define $f^∞_{(t,
\obstime)}$ in terms of morphisms $f^∞_{(t′, \obstime)}$ with $t′ > t$. So if
the order~$≥$ on the set $\set{t ∈ T}{t ≤ \obstime}$ is well-founded,
\eqref{equation:f-infinity-element-definition}
and~\eqref{equation:f-infinity-element-definition-helper} define $f^∞_{(t,
\obstime)}$ by well-founded recursion with recursion parameter~$t$. As a
consequence, there exists a unique $f^∞$ that fulfills
\eqref{equation:f-infinity-for-apc}.

Based on the above considerations, we require for an ℛ-CPC that for all $t ∈ T$,
the order~$≥$ on $\set{t′ ∈ T}{t′ ≤ t}$ is well-founded, which means that we
cannot have an infinite ascending sequence of times that all lie before a
certain time. This makes $ω$-supertasks~\cite{laraudogoitia:supertasks}
impossible, which, for example, implies that the sequence of actions described
in Zeno’s paradox of Achilles and the tortoise cannot occur.

Note however that we still allow time scales that are quite different from the
discrete time scale $(ℕ, ≤)$. For example, $\bigset{z + 1/n}{z ∈ ℤ \mathrel∧ n ∈
ℕ ∖ \{0\}}$ with the usual ordering of rational numbers is still a perfectly
acceptable time scale.

\extsubsection{enabling-recursion-on-processes}{Enabling Recursion on Processes}

The APC $\left(𝒲, ℬ^ℐ, ▷″, θ″, ϑ″\right)$ induced by a CPC $(T, ≤, ℬ)$ models
recursion on processes if and only if for every morphism $f : (A × C)
▷″_{\timebound} B → C$, there exists a unique morphism~$f^*$ for which the
following holds:
\begin{align}
f^* & \relwithsizeof=: A ▷″_{\timebound} B → C                                  \notag\\
\label{equation:f-star-for-apc}
f^* & =                f                                                      ∘
                       \left((\id_A × f^*) ▷″_{\id_{\timebound}} \id_B\right) ∘
                       θ″_{\timebound, A, B}
\end{align}
Equation~\eqref{equation:f-star-for-apc} is true if and only if for all $t,
\obstime ∈ T$ with $t ≤ \obstime$, we have
\begin{equation}
\label{equation:f-star-element-definition}
f^*_{(t, \obstime)} = f_{(t, \obstime)}                                  ∘
                      h                                                  ∘
                      \left(θ″_{\timebound, A, B}\right)_{(t, \obstime)}
\end{equation}
where $h$ is defined as follows:
\begin{align}
h      & = \begin{cases}
           \id_0                                                                                          &
               \text {if $\timebound < t$}                                                                \\
           ∐_{t′ ∈ (t, \timebound]} s_{t′}                                                                &
               \text {if $t ≤ \timebound ≤ \obstime$}                                                     \\
           ∐_{t′ ∈ (t, \obstime]} s_{t′} + ∏_{t′ ∈ (t, \obstime]} \left(\id × f^*_{(t′, \obstime)}\right) &
               \text {if $\obstime <_∞ \timebound$}
           \end{cases}                                                                                       \\
\label{equation:last-f-star-element-definition-helper}
s_{t′} & = \left(∏_{t″ ∈ (t, t′)} \left(\id × f^*_{(t″, \obstime)}\right)\right) × \id
\end{align}
Equations \eqref{equation:f-star-element-definition}
through~\eqref{equation:last-f-star-element-definition-helper} define $f^*_{(t,
\obstime)}$ by well-founded recursion, assuming that the order~$≥$ on the set
$\set{t ∈ T}{t ≤ \obstime}$ is well-founded. As a consequence, there exists a
unique $f^*$ that fulfills \eqref{equation:f-star-for-apc}.

\extsubsection{cpcs-with-recursion-and-corecursion-summary}{Summary}

We formulate the definition of ℛ-CPCs based on the above explanations.

\begin{extdefinition}{cpc-with-recursion-and-corecursion}
                     [CPC with recursion and corecursion]
A CPC $(T, ≤, ℬ)$ is a CPC with recursion and corecursion (ℛ-CPC) if and only if
for all $t ∈ T$, the order~$≥$ on $\set{t′ ∈ T}{t′ ≤ t}$ is well-founded.
\end{extdefinition}

We have developed ℛ-CPCs such that they give rise to ℛ-APCs. We state the
relationship between ℛ-CPCs and ℛ-APCs in the following theorem.

\begin{exttheorem}{r-apc-from-r-cpc}
The APC that is induced by an ℛ-CPC according to \theoremref{apc-from-cpc} is an
ℛ-APC.
\end{exttheorem}

\extsection{related-work}{Related Work}

We can build ℛ-APCs by taking the categorical semantics of the intuitionistic
modal logic IK~\cite{bellin:m4m-2001} and successively adding more structure.
This leads us first to intuitionistic S4 categories
\cite[Section~4]{jeltsch:entcs-286}, which are closely related to the
categorical semantics of intuitionistic S4 variants by
Kobayashi~\cite{kobayashi:tcs-175-1} and Bierman and
de~Paiva~\cite{bierman:sl-65-3}. Subsequent additions to intuitionistic S4
categories lead us to temporal categories \cite[Sections 5
and~6]{jeltsch:entcs-286}, from there to APCs~\cite{jeltsch:plpv-2014}, and
finally to ℛ-APCs.

Krishnaswami and Benton~\cite{krishnaswami:lics-2011} model FRP categorically
based on 1-bounded ultrametric spaces. Their semantics ensures that morphisms
only model causal functions, which is also a key property of CPCs. However they
model only behaviors, not events or even processes. Furthermore they require the
time scale to be discrete. Although this is a restriction, it makes it possible
to model recursion on behaviors.

Cave et~al.~\cite{cave:popl-2014} present an FRP calculus that features
inductive and coinductive types and a type constructor that corresponds to the
“next” modality of LTL. The authors show that process types are definable within
their calculus. We strongly conjecture that their calculus also allows for the
definition of recursion and corecursion operators on processes. The downside of
their solution is that it is fundamentally tied to discrete time.

Jeffrey \cite[Section~2]{jeffrey:plpv-2012} defines categories ${⊟}\rset$ and
${⊵}\rset$. The morphisms of ${⊟}\rset$ model causal functions on time-varying
values, and the morphisms of ${⊵}\rset$ model causal functions that produce
output values based only on a finite history. Jeffrey also introduces a fixed
point operator $\ifix$, which is similar to our ${-}^*$ operator for recursion,
but uses time-varying values that lie in the past, while ${-}^*$ uses processes,
whose continuous parts lie in the future. Consequently Jeffrey requires that
$≤$, and not $≥$, is well-founded on certain intervals.

Birkedal et~al.~\cite{birkedal:lmcs-8-4} model guarded recursion by the topos of
trees. The topos of trees is the category $𝒮 = \Set^{\op{𝒩}}$, where $𝒩$ is the
category of the ordered set $(ℕ, ≤)$. If $A ∈ \Ob(𝒮)$ models a type~$τ$, an
object~$An$ with $n ∈ ℕ$ models the type whose inhabitants give the information
we have about inhabitants of~$τ$ after $n$ recursion steps, and a morphism~$Av$
with $v : m → n$ models the function that forgets the information acquired
during recursion steps $n + 1$ through~$m$. There is a close connection between
this development and CPCs. Consider the CPC $(ℕ, ≤, \Set)$. Let $ℐ$ be the
temporal index category of this CPC, and let $ℰ$ be the full subcategory
of~$\Set^ℐ$ that is induced by those objects that model types with
time-independent inhabitance, that is, by those $A ∈ \Ob\left(\Set^ℐ\right)$
with $∀ k, l, n ∈ ℕ . A(k, k + n) = A(l, l + n)$ and a similar condition for
morphisms. Then the categories $𝒮$ and~$ℰ$ are isomorphic. Birkedal et~al.
define a fixed point operator based on a “later” type constructor that
corresponds to the “next” modality of temporal logic. Studying the relationship
between this fixed point operator and our structure for recursion and
corecursion on processes remains a task for the future.

\extsection{conclusions-and-further-work}{Conclusions and Further Work}

We have developed ℛ-APCs and ℛ-CPCs, which are categorical models of FRP with
recursion and corecursion on processes. ℛ-APCs are defined purely axiomatically,
while ℛ-CPCs use concrete structure to express time-dependence of type
inhabitance and causality of FRP operations. We have furthermore shown that
ℛ-APCs generalize ℛ-CPCs.

In the future, we want to develop categorical models of FRP with support for
mutable state. Another goal of ours is to implement FRP in mainstream functional
programming languages such that the interface directly reflects the abstract
categorical semantics.

\extappendix{changes-in-the-completely-iterative-monad-definition}
            {Changes in the Completely Iterative Monad Definition}

Milius \cite[Definition~5.5]{milius:ic-196-1} defines completely iterative
monads differently from us. A simplified version of his definition is as
follows.

\begin{extdefinition}{milius-style-completely-iterative-monad}
                     [Milius-style completely iterative monad]
Let $𝒞$ be a category with binary coproducts. A pair $(T′, μ′)$ is a
Milius-style completely iterative monad on~$𝒞$ if and only if it is an ideal
monad on~$𝒞$, and for every morphism $f : C → B + T′(B + C)$, there exists a
unique morphism $f^† : C → TB$ with $T = \Id + T′$ for which the diagram in
\figureref{condition-for-a-morphism-f-dagger-of-a-milius-style-completely-iterative-monad}
commutes.
\end{extdefinition}
\begin{extfigure}{condition-for-a-morphism-f-dagger-of-a-milius-style-completely-iterative-monad}
                 {Condition for a morphism~$f^†$ of a Milius-style completely iterative monad}

\begin{tikzpicture}[diagram,column sep=7em]

\matrix{
\node [object] (start)            {$C$};             &
\node [object] (end)              {$TB$};            \\
\node [object] (after expansion)  {$B + T′(B + C)$}; &
\node [object] (after inner call) {$B + T′TB$};      \\
};

\edges

\path (start)            edge [morphism] node [auto=left]  {$f^†$}                            (end)
      (start)            edge [morphism] node [auto=right] {$f$}                              (after expansion)
      (after expansion)  edge [morphism] node [auto=right] {$\id_B + T′\left[ι₁, f^†\right]$} (after inner call)
      (after inner call) edge [morphism] node [auto=right] {$\id_B + μ′_B$}                   (end);

\end{tikzpicture}

\end{extfigure}

We do not use this definition in this paper, because in our opinion, it has
disadvantages:
\begin{itemize}

\item

The morphism~$f : C → B + T′(B + C)$ gives us two points per iteration where
iterating can come to an end. These two points correspond to the two sums in the
codomain of~$f$. However only one such point is necessary.

\item

The codomain of~$f^†$ uses the derived functor~$T$ instead of the actual ideal
monad functor~$T′$, weakening the type of~$f^†$.

\end{itemize}
However our own definition is equivalent to the one by Milius. This fact is
expressed by the following two theorems.

\begin{exttheorem}{jeltsch-cim-is-milius-cim}
If $(T′, μ′)$ is a completely iterative monad on a category~$𝒞$ according to
\definitionref{completely-iterative-monad}, then it is a Milius-style completely
iterative monad on~$𝒞$.
\end{exttheorem}

\begin{extproof}
Let $(T′, μ′)$ be a completely iterative monad according to
\definitionref{completely-iterative-monad}, and let $g$ be an arbitrary morphism
with $g : C → B + T′(B + C)$. We have to show that there exists a unique
morphism~$g^†$ for which the following holds:
\begin{align}
g^† & \relwithsizeof=: C → B + T′B                                                         \notag\\
g^† & =                \left(\id_B + \left(μ′_B ∘ T′\left[ι₁, g^†\right]\right)\right) ∘ g
\end{align}
We define $f$ as follows:
\begin{align}
f & \relwithsizeof=: T′(B + C) → T′(B + T′(B + C)) \notag\\
\label{equation:f-in-equivalence-proof-1}
f & =                T′[ι₁, g]
\end{align}
Since $(T′, μ')$ is a completely iterative monad according to
\definitionref{completely-iterative-monad}, there exists a unique $f^∞$ for
which the following holds:
\begin{align}
f^∞ & \relwithsizeof=: T′(B + C) → T′B            \notag\\
\label{equation:f-infinity-in-equivalence-proof-1}
f^∞ & =                μ′_B ∘ T′(\id_B + f^∞) ∘ f
\end{align}
According to \eqref{equation:f-in-equivalence-proof-1}, we have
\begin{equation}
\begin{split}
μ′_B ∘ T′(\id_B + f^∞) ∘ f & = μ′_B ∘ T′(\id_B + f^∞) ∘ T′[ι₁, g]        \\
                           & = μ′_B ∘ T′[ι₁, (\id_B + f^∞) ∘ g]\enspace,
\end{split}
\end{equation}
so \eqref{equation:f-infinity-in-equivalence-proof-1} is equivalent to the
following equation:
\begin{equation}
f^∞ = μ′_B ∘ T′[ι₁, (\id_B + f^∞) ∘ g]
\end{equation}
We define the functions $q$ and~$r$ on morphisms as follows:
\begin{align}
& q \relwithsizeof=: \Hom(T′(B + C), T′B) → \Hom(C, B + T′B) \notag\\
& q(a) =             (\id_B + a) ∘ g                      \\[\topsep]
& r \relwithsizeof=: \Hom(C, B + T′B) → \Hom(T′(B + C), T′B) \notag\\
& r(b) =             μ′_B ∘ T′[ι₁, b]
\end{align}
We know that there exists a unique $f^∞$ with
\begin{equation}
\label{equation:f-infinity-simplified-in-equivalence-proof-1}
f^∞ = r(q(f^∞))\enspace,
\end{equation}
and we have to prove that there exists a unique $g^†$ with
\begin{equation}
\label{equation:g-dagger-simplified-in-equivalence-proof-1}
g^† = q\left(r\left(g^†\right)\right)\enspace.
\end{equation}
From \eqref{equation:f-infinity-simplified-in-equivalence-proof-1}, it follows
that
\begin{equation}
q(f^∞) = q(r(q(f^∞)))\enspace,
\end{equation}
so $g^† = q(f^∞)$ fulfills
\eqref{equation:g-dagger-simplified-in-equivalence-proof-1}. On the other hand,
if we have a $g^†$ that fulfills
\eqref{equation:g-dagger-simplified-in-equivalence-proof-1}, we know that
\begin{equation}
r\left(g^†\right) = r\left(q\left(r\left(g^†\right)\right)\right)\enspace.
\end{equation}
Because there is only one $f^∞$ that fulfills
\eqref{equation:f-infinity-simplified-in-equivalence-proof-1}, it follows that
$f^∞ = r\left(g^†\right)$. From this, we get
\begin{equation}
g^† = q\left(r\left(g^†\right)\right) = q(f^∞)\enspace.
\end{equation}
So $g^† = q(f^∞)$ is the unique solution of
\eqref{equation:g-dagger-simplified-in-equivalence-proof-1}.
\end{extproof}

\begin{exttheorem}{milius-cim-is-jeltsch-cim}
If $(T′, μ′)$ is a Milius-style completely iterative monad on a category~$𝒞$,
then it is a completely iterative monad on~$𝒞$ according to
\definitionref{completely-iterative-monad}.
\end{exttheorem}

\begin{extproof}
Let $(T′, μ′)$ be a Milius-style completely iterative monad, and let $f$ be an
arbitrary morphism with $f : C → T′(B + C)$. We have to show that there exists a
unique morphism~$f^∞$ for which the following holds:
\begin{align}
f^∞ & \relwithsizeof=: C → T′B                     \notag\\
f^∞ & =                μ′_B ∘ T′[ι₁, ι₂ ∘ f^∞] ∘ f
\end{align}
We define $g$ as follows:
\begin{align}
g & \relwithsizeof=: C → B + T′(B + C) \notag\\
\label{equation:g-in-equivalence-proof-2}
g & =                ι₂ ∘ f
\end{align}
Since $(T′, μ')$ is a Milius-style completely iterative monad, there exists a
unique $g^†$ for which the following holds:
\begin{align}
g^† & \relwithsizeof=: C → B + T′B                                                         \notag\\
\label{equation:g-dagger-in-equivalence-proof-2}
g^† & =                \left(\id_B + \left(μ′_B ∘ T′\left[ι₁, g^†\right]\right)\right) ∘ g
\end{align}
According to \eqref{equation:g-in-equivalence-proof-2}, we have
\begin{equation}
\begin{split}
\left(\id_B + \left(μ′_B ∘ T′\left[ι₁, g^†\right]\right)\right) ∘ g
    & = \left(\id_B + \left(μ′_B ∘ T′\left[ι₁, g^†\right]\right)\right) ∘ ι₂ ∘ f \\
    & = ι₂ ∘ μ′_B ∘ T′\left[ι₁, g^†\right] ∘ f\enspace,
\end{split}
\end{equation}
so \eqref{equation:g-dagger-in-equivalence-proof-2} is equivalent to the
following equation:
\begin{equation}
g^† = ι₂ ∘ μ′_B ∘ T′\left[ι₁, g^†\right] ∘ f
\end{equation}
We define the functions $q$ and~$r$ on morphisms as follows:
\begin{align}
& q \relwithsizeof=: \Hom(C, B + T′B) → \Hom(C, T′B) \notag\\
& q(a) =             μ′_B ∘ T′[ι₁, a] ∘ f            \\[\topsep]
& r \relwithsizeof=: \Hom(C, T′B) → \Hom(C, B + T′B) \notag\\
& r(b) =             ι₂ ∘ b
\end{align}
We know that there exists a unique $g^†$ with
\begin{equation}
g^† = r\left(q\left(g^†\right)\right)\enspace,
\end{equation}
and we have to prove that there exists a unique $f^∞$ with
\begin{equation}
\label{equation:f-infinity-simplified-in-equivalence-proof-2}
f^∞ = q(r(f^∞))\enspace.
\end{equation}
Using reasoning analogous to the one at the end of the previous proof, we can
deduce that $f^∞ = q\left(g^†\right)$ is the unique solution of
\eqref{equation:f-infinity-simplified-in-equivalence-proof-2}.
\end{extproof}

\section*{Acknowledgments}

I thank Tarmo Uustalu for all the helpful discussions about the topics of this
paper, and in particular, for introducing me to completely iterative monads and
recursive comonads.

This work was supported by the target-financed research theme No.~0140007s12 of
the Estonian Ministry of Education and Research and by the ERDF through the
Estonian Center of Excellence in Computer Science (EXCS) and the national ICTP
project \emph{Coinduction for Semantics, Analysis, and Verification of
Communicating and Concurrent Reactive Software}. I thank all tax payers in the
European Union for funding the ERDF.

\bibliographystyle{eptcs}
\bibliography{references/references}

\end{document}